\documentclass[prl,aps,twocolumn,superscriptaddress,showpacs,floatfix,preprintnumbers]{revtex4-2}
\usepackage{amsmath}
\usepackage{mathtools}
\usepackage{graphicx} % Required for inserting images
\usepackage{tikz}
\usetikzlibrary{arrows}

\usepackage[colorlinks=true,urlcolor=blue,linkcolor=blue,citecolor=blue]{hyperref}
\usepackage[normalem]{ulem}

\usepackage[export]{adjustbox}
 
\usepackage{comment}
\usepackage{flushend}

\usepackage[T1]{fontenc} % if needed

\usepackage{mathtools}

\usepackage{xcolor}
\usepackage{graphicx}% Include figure files
\usepackage{dcolumn}% Align table columns on decimal point
\usepackage{bm}% bold math
\usepackage{multirow}

\usepackage{epstopdf}
\usepackage[capitalize]{cleveref}

\newcommand{\Beq}{\begin{equation}\begin{aligned}}
\newcommand{\Eeq}{\end{aligned}\end{equation}}

\begin{document}

\preprint{YITP-26-107}

\title{Bumpy inflation from explosive particle production} 

\author{Kaloian D. Lozanov}
\email{kaloyan.lozanov@tu-sofia.bg}
\affiliation{Department of Applied Physics, Faculty of Applied Mathematics and Informatics,
Technical University of Sofia,
8, Saint Kliment Ohridski Blvd, Sofia 1000, Bulgaria.}
\affiliation{Asia Pacific Center for Theoretical Physics (APCTP), Pohang 37673, Republic of Korea.}
\affiliation{Kavli Institute for the Physics and Mathematics of the Universe (WPI), UTIAS
The University of Tokyo, Kashiwa, Chiba 277-8583, Japan.}
\author{Xavier Pritchard}
\email{x.pritchard@sussex.ac.uk}
\affiliation{Department of Physics and Astronomy, University of Sussex, Brighton BN1 9QH, UK.}
\author{Misao Sasaki}
\email{misao.sasaki@apctp.org}
\affiliation{Asia Pacific Center for Theoretical Physics (APCTP), Pohang 37673, Republic of Korea.}
\affiliation{Kavli Institute for the Physics and Mathematics of the Universe (WPI), UTIAS
The University of Tokyo, Kashiwa, Chiba 277-8583, Japan.}
\affiliation{Center for Gravitational Physics and Quantum Information, Yukawa Institute for Theoretical Physics,
Kyoto University, Kyoto 606-8502, Japan.}
\affiliation{Leung Center for Cosmology and Particle Astrophysics, National Taiwan
University, Taipei 10617, Taiwan.}
\affiliation{Department of Physics, Pohang University of Science and Technology (POSTECH),\\ Pohang 37673, Republic of Korea}

\date{\today}

\begin{abstract}
We present a first-principles derivation of an effective inflaton potential with a bumpy feature.
We consider the coupling of a spectator field $\chi$ to the inflaton $\phi$ such that 
$\chi$ is massive enough for almost all values of $\phi$, but becomes massless when the inflaton crossed the value $\phi_\star$. 
This gives rise to a burst of $\chi$-particle production at $\phi=\phi_\star$. By taking the backreaction of the particle production, we obtain a bumpy feature in the effective potential.
%We study an effectively single-field model of slow-roll inflation, in which the inflaton is coupled to a sub-dominant spectator scalar field.
%with a vanishing bare mass. 
%The interaction generates a non-zero inflaton-dependent mass for the spectator field. We choose the interaction, so that it vanishes briefly for a specific value of the inflaton during which the spectator scalar becomes massless. 
%This leads to non-adiabatic spectator-field particle production.
We discover that for certain choices of the interaction constants and field scale relevant for particle production, 
%the backreaction from the spectator field on the inflaton can give rise to bump-like feature in the inflaton potential. 
the backreaction-induced bumpy effective potential can lead to a sizable enhancement in the curvature perturbation. In particular, the enhancement can be large enough to generate observationally relevant amounts of primordial black holes and stochastic gravitational wave backgrounds.
\end{abstract}

\maketitle

{\it Introduction}.--The paradigm of cosmic inflation \cite{BROUT197878,Starobinsky:1979ty,Sato:1981qmu,PhysRevD.23.347,LINDE1983177} provides an explanation for multiple challenging questions posed by standard cosmology such as the horizon and flatness problems.  The inflaton background drives the FLRW evolution of the universe during inflation, which takes the form of quasi-de-Sitter expansion. The quantized perturbations in the inflaton lead to the small departures from FLRW spacetime \cite{Mukhanov:1981xt,Sasaki:1986hm,Kodama:1984ziu} observed in the Cosmic Microwave Background (CMB) \cite{Planck:2018jri,Planck:2018vyg}. 
 The perturbations on smaller, sub-CMB, scales can be enhanced due to inflaton self-interactions \cite{Abolhasani:2019cqw} and can lead to the generation \cite{Shibata:1999zs} of PBHs after inflation. 
The PBHs can play the role of dark matter \cite{Sasaki:2016jop,Sasaki:2018dmp}, and the density variations giving rise to their formation can source scalar-induced gravitational waves (SIGWs) \cite{Domenech:2021ztg}.

Such enhancements can arise from features in the inflaton self-interaction potential, including, for example, a plateau-like flat region \cite{Kinney:2005vj,Dimopoulos:2017ged}, a bump \cite{Cai:2022erk}, a sharp step-like descent \cite{Kristiano:2024vst}, a series of bumps and troughs \cite{Inomata:2022yte}. 
In all these works, however, the existence of features in the inflaton potential is simply assumed in an ad hoc manner. Therefore, it is intriguing if one can derive such features from first principles.  

In the current work, we derive for the first time a bumpy feature in the inflaton potential by going beyond single-field inflation. 
We consider a scenario in which the inflaton $\phi$ interacts with a sub-dominant daughter field -- a spectator field $\chi$. 
We choose an interaction such that $\chi$ is mostly heavy enough to remain unexcited, but it becomes massless at the moment when $\phi$ crosses a certain value, $\phi=\phi_\star$. 
This leads to a burst-like production of spectator particles.

Particle production during inflation can arise through a variety of non-adiabatic mechanisms. Classic examples include parametric resonances \cite{Kofman:1997yn, Greene:1998nh,Chung:1999ve,Gorgulho:2025wxz} and particle production associated with phase transitions \cite{1990PhRvD..42.2491T,Linde:1993cn,Felder:2000hj,Felder:2001kt}. A distinct mechanism is instant preheating, in which a rapid passage of the inflaton through a point where a coupled field becomes light leads to efficient non-adiabatic particle production without requiring sustained inflaton oscillations or parametric resonance \cite{Felder:1998vq,Felder:1999pv,Allahverdi:2010xz}. 

The effect of such an event on the background and primordial perturbations, has been studied extensively \cite{2003JCAP...09..008E,Kofman:2004yc,Barnaby:2009dd,Barnaby:2009mc,Fedderke:2014ura,Pearce:2017bdc,Goolsby-Cole:2017hod,Cai:2021yvq,Yu:2023ity}. In these studies, however, the particle production event is treated as a source for fluctuations; the bump in the curvature power spectrum comes from the rescattering of the produced particles off the inflaton condensate. 

Rather than treating the particle production primarily as a source of additional inflaton fluctuations, we instead posit that this backreaction may be interpreted as a local feature in the effective inflaton potential. 
Taking into account the backreaction of the produced particles on the inflaton dynamics in a self-consistent manner, we then study the corresponding impact this has on the curvature power spectrum.
%We take into account the backreaction of the produced particles on the inflaton dynamics in a self-consistent manner. 
%We show that this gives rise to an effectively single-field dynamics with an effective inflaton potential having a bump-like feature.
We demonstrate that the bump can cause a significant enhancement of the curvature perturbation, potentially reaching amplitudes relevant for the generation of PBHs and associated scalar-induced gravitational waves.

%The paper is organized as follows. In Section \ref{sec:model} we present the model of inflation. We outline in Section \ref{sec:backdyn} the background dynamics of the inflaton field, driving the quasi-de-Sitter stage of inflationary expansion. We study in Section \ref{sec:partprod} the production of daughter particles in the inflationary background. In Section \ref{sec:effinflpot} we show that the corrections in the inflaton potential due to backreaction from the daughter particles take the form of a bump. We determine in Section \ref{sec:curvenh} that the bump-like feature in the effective single-field inflaton potential can give rise to a sizable enhancement in the curvature perturbation during inflation. We give our concluding remarks in Section \ref{sec:concl}. Technical details are relegated to the Appendices. 

Throughout the paper we work in natural units in which $\hbar=c=1$ and the reduced Planck mass is $m_{pl}=\sqrt{1/(8\pi G)}=2.435\times10^{18}\,\rm{GeV}$.

\begin{figure}[th!]
\includegraphics[width=3.3in]{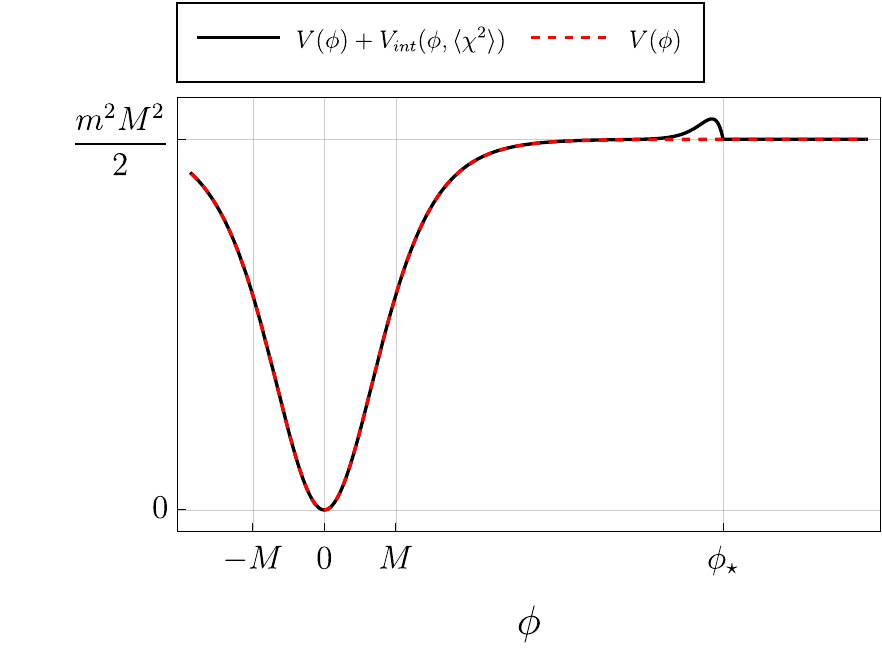}
\caption{The bare inflaton potential in dashed red, given by the $\alpha$-attractor expression in eq.~\eqref{eq:V}, and the effective inflaton potential in black, given in eqs.~\eqref{eq:Veff} and \eqref{eq:Vintf}. A bump like feature in the effective potential is induced by the interaction between the inflaton and the spectator field. 
The interaction leads to the generation of daughter particles when $\phi=\phi_{\star}$ and to their backreaction when $\phi<\phi_{\star}$. 
At early times, $\phi>\phi_{\star}$, the interaction potential has no effect and slow-roll inflation proceeds in the standard fashion.}
    \label{fig:papV}
\end{figure}                     

{\it The model}--We consider a two-field model of inflation, 
\Beq
S=\int d^4x\sqrt{-g}\Big[&\frac{m_{pl}^2}{2}R-\frac{(\partial\phi)^2}{2}-V(\phi)\\
&-\frac{(\partial\chi)^2}{2}-V_{int}(\phi,\chi^2)\Big]\,,
\Eeq
where $\phi$ plays the role of the inflaton field and $\chi$ the daughter field. 
For concreteness, we assume the inflaton potential is given by the $\alpha$-attractor form,
\Beq
\label{eq:V}
V(\phi)=\frac{m^2M^2}{2}\tanh^2\left(\frac{\phi}{M}\right)\,.
\Eeq
The interaction potential is assumed to have the form,
\Beq
\label{eq:Vint}
V_{int}(\phi,\chi^2)=\frac{\lambda}{2}\left(\phi-\phi_{\star}\right)^2\chi^2\,,
\Eeq
where $\lambda$ should be large enough ($\lambda\gg1$) so that we obtain an explosive particle production in a very short period of time. 
We shall see later that our results are consistent with this condition. 

The inflaton potential from eq.~\eqref{eq:V} can give rise to successful single-field inflation -- $\alpha$-attractor inflation -- consistent with CMB observations \cite{Kallosh:2013pby,Kallosh:2013lkr,Kallosh:2013hoa}. 
In particular, it predicts observationally consistent values for the measured spectral index, $n_s$, and the constrained tensor-to-scalar ratio, $r_{0.002}$, given by
\Beq
n_s\approx 1-\frac{2}{N_{\rm CMB}}\,,~
r_{0.002}\approx\frac{2}{N_{\rm CMB}^2}\frac{M^2}{m_{pl}^2}\,,
\Eeq
provided $M\lesssim m_{pl}$ and the number of {\it e}-folds of inflation when the CMB scale leaves the horizon is $N_{\rm CMB}\approx60$. The measured amplitude of scalar power, $A_s\approx 2.2\times10^{-9}$, is attained if
\begin{equation}
m^2\approx \frac{6\pi^2 A_s}{N_{\rm CMB}^2}m_{pl}^2\,,
\end{equation}
where $N_{\rm CMB}$ is the number of $e$-folds until the end of inflation if there were no effect of particle production. 
The interaction potential from eq.~\eqref{eq:Vint} leads to a sudden burst of $\chi$-particle production near $\phi\approx\phi_{\star}$.
Since this slows down the inflaton motion, the actual number of $e$-folds would be slightly larger than $N_{\rm CMB}$.

%The interaction potential from eq.~\eqref{eq:Vint} leads to a sudden burst of $\chi$-particle production near $\phi\approx\phi_{\star}$.
In the rest of this work, we study the backreaction of the particle production, which leads to an effective inflaton potential,
\Beq
\label{eq:Veff}
V_{eff}(\phi)=V(\phi)+V_{int}(\phi,\langle\hat\chi^2\rangle(\phi))\,,
\Eeq
where $\langle\hat\chi^2\rangle(\phi)$ is the expectation value determined by the details of the particle production. 
%Since $V_{int}>0$, the backreaction leads to a bump-like modification of the inflaton potential for values of $\phi$ slightly smaller than $\phi_{\star}$
%
In our scenario, the particle production causes $\langle\hat\chi^2\rangle$ to become non-zero for $\phi<\phi_\star$, and the resulting positive contribution to $V_{int}(\phi,\langle\hat\chi^2\rangle(\phi))$ produces a localised bump-like modification to the inflaton potential, see Fig. \ref{fig:papV}. 

Particle production backreaction on the inflaton dynamics is closely related to trapped inflation \cite{Green:2009ds,Lee:2011fj,Furuta:2025fbh}, where the repeated production of particles can slow the inflaton and modify its subsequent evolution. In contrast, here we consider a single localized production event, whose backreaction gives rise to a transient, localized modification of the effective inflaton dynamics.
In the following sections, we investigate how this bump forms and what its consequences are for the power spectrum of the curvature perturbation.

{\it Background dynamics}.--We begin with the background dynamics of the inflaton field and FLRW spacetime metric. In particular, at the background level we have
\Beq
\phi(t,x^i)=\bar{\phi}(t)
\Eeq
and
\Beq
g_{\mu\nu}(t,x)&=\bar{g}_{\mu\nu}(t)\\
&={\rm diag}[-1,a^2(t),a^2(t),a^2(t)]\,,
\Eeq
where we assume spatially-flat FLRW spacetime. The Klein-Gordon equation of motion \cite{Gordon:1926emj,Klein:1926tv,Schrodinger:1926iou} and the Hamiltonian constraint take the standard forms, respectively,
\Beq
\label{eq:KGeom}
\ddot{\bar{\phi}}+3H\dot{\bar{\phi}}+\partial_{\bar{\phi}}V_{eff}=0\,,
\Eeq
and
\Beq
\label{eq:Freom}
H\equiv\frac{\dot{a}}{a}=\sqrt{\frac{1}{3m_{pl}^2}\left(\frac{\dot{\bar{\phi}}}{2}+V_{eff}(\bar{\phi})\right)}\,.
\Eeq

We also recall the expression for the first slow-roll parameter,
\Beq
\epsilon\equiv-\frac{\dot{H}}{H^2}=\frac{\dot{\bar{\phi}}^2}{2m_{pl}^2H^2}\,.
\Eeq

\if0
Since when $\bar{\phi}>\phi_{\star}:\quad V_{eff}(\bar{\phi})=V(\bar{\phi})$, in this field range inflation proceeds in the usual fashion, following the $\alpha$-attractor slow-roll solution. Later on, at $\phi=\phi_{\star}$ abrupt $\chi$-particle production takes place. Subsequently, when $\bar{\phi}<\phi_{\star}:\quad V_{eff}(\bar{\phi})\neq V(\bar{\phi})$, due to the backreaction-induced effective bump in the inflaton potential. The bump slows down the slow-roll of the inflaton. We assume for the inflaton field velocity when $\bar{\phi}(t)\approx\phi_{\star}$ the following approximate step-like evolution
\Beq
\dot{\bar{\phi}}(t)=\dot{\bar{\phi}}_{\star,-}\Theta(t_\star-t)+\dot{\bar{\phi}}_{\star,+}\Theta(t-t_\star)\,,
\Eeq
where $\bar{\phi}(t_\star)=\phi_{\star}$. We characterize the slowing down of the inflaton slow-roll with the ratio parameter
\Beq
g\equiv\frac{\left|\dot{\bar{\phi}}_{\star,+}\right|}{\left|\dot{\bar{\phi}}_{\star,-}\right|}\,,
\Eeq
where $g\leq1$.
\fi

{\it Particle production}.--To study the production of $\chi$ particles we consider the $\chi$ field as a quantized perturbation on top of the classical inflaton and spacetime metric backgrounds. In particular, we define the field (with a canonical kinetic term) $\hat{X}(t,x^i)\equiv a^{3/2}(t)\hat{\chi}(t,x^i)$. Its Fourier modes are quantized in the usual manner
\begin{equation}
    \begin{aligned}
        &\hat{X}(t,x^i)=\int \frac{d^3k}{(2\pi)^3}\left[e^{ik^jx^j}X_k(t)\hat{b}_k+h.c.\right]\,,
    \end{aligned}
\end{equation}
where the time-independent $\hat{b}_k$ and $\hat{b}_k^\dagger$ operators play the role of annihilation and creation operators, respectively, and obey the commutation relation $[\hat{b}_k,\hat{b}_{k'}^\dagger]=\delta(k-k')$. The mode function obeys the classical equation of motion
\Beq
\ddot{X}_k+\left(\frac{k^2}{a^2}+\lambda(\phi-\phi_{\star})^2-\frac{3}{2}\dot{H}-\frac{9}{4}H^2\right)X_k=0\,,
\Eeq
We consider the excitation of modes which are sub-horizon during slow-roll inflation, $(k/a)^2\gg H^2\gg|\dot{H}|$ for which 
\Beq
\label{eq:Xeom}
\ddot{X}_k+\left(\frac{k^2}{a^2}+\lambda(\phi-\phi_{\star})^2\right)X_k=0\,,
\Eeq
holds. We will see that these sub-horizon modes give the dominant backreaction.

To study the particle production we consider eq. \eqref{eq:Xeom} near $\phi\approx\phi_{\star}$.
As we expect that a bursting particle production leads to a bumpy feature in the effective potential, we approximate the inflaton dynamics near $\phi_{\star}$ by as 
\Beq
{\bar{\phi}}(t)-\phi_{\star}=\left(\dot{\bar{\phi}}_{\star,-}\Theta(t_\star-t)+\dot{\bar{\phi}}_{\star,+}\Theta(t-t_\star)\right)(t_\star-t)\,,
\Eeq
and set $a(T)\approx a(t_\star)\equiv a_\star$. 
Here the value of $\dot{\bar{\phi}}_{\star,-}$ is determined by the slow-roll evolution and that of  $\dot{\bar{\phi}}_{\star,+}$ is determined self-consistently by taking account of the backreaction due to the particle production.
Then for $t<t_\star$, eq. \eqref{eq:Xeom} reduces to
\Beq
\label{eq:Xmeom}
\frac{d^2}{dz_-^2}X_{\kappa_-}(z_-)+\left(\kappa_-^2+z_-^2\right)X_{\kappa_-}(z_-)=0\,,
\Eeq
and for $t>t_\star$, eq. \eqref{eq:Xeom} reduces to
\Beq
\label{eq:Xpleom}
\frac{d^2}{dz_+^2}X_{\kappa_+}(z_+)+\left(\kappa_+^2+z_+^2\right)X_{\kappa_+}(z_+)=0\,,
\Eeq
where
\Beq
\kappa_{\mp}&\equiv\frac{k}{a_\star\lambda^{1/4}\left|\dot{\bar{\phi}}_{\star,\mp}\right|^{1/2}}\,,\\
z_{\mp}&\equiv \lambda^{1/4}\left|\dot{\bar{\phi}}_{\star,\mp}\right|^{1/2}(t-t_\star)\,.
\Eeq
Note that $\kappa_-(k)=\kappa_+(k)g^{1/2}$, where $g=|\dot{\bar{\phi}}_{\star,+}/\dot{\bar{\phi}}_{\star,-}|$ is the ratio of the inflaton velocity before the feature, versus at the peak. 

The effective time-dependent frequencies in %the harmonic-oscillator like
eqs.~\eqref{eq:Xmeom} and \eqref{eq:Xpleom} are
\Beq
\omega^2_\mp=\kappa_\mp^2+z_\mp^2\,.
\Eeq
We note that the time-dependence of the frequency is adiabatic,
\Beq
\left|\frac{\omega'(z_\mp)}{\omega_\mp^2(z_\mp)}\right|\ll1\,,
\Eeq
unless $\kappa_\mp\rightarrow0$ and $z_\mp\rightarrow\mp0$, in which case there is a violation of adiabaticity,
\Beq
\label{eq:nonad}
\left|\frac{\omega'(z_\mp)}{\omega_\mp^2}\right|\sim1\,.
\Eeq
We show below that this leads to non-adiabatic particle production.

Eqs. \eqref{eq:Xmeom} and \eqref{eq:Xpleom} are of the form of the Weber equation, and their solutions are given in terms of parabolic cylinder functions \cite{Kofman:1997yn},
\Beq
\label{eq:Xparcylfncns}
X_{\kappa_\mp}(z_\mp)=C_{1,\mp}&D_{-\frac{1}{2}-i\frac{\kappa_\mp^2}{2}}((1+i)z_{\mp})\\
&+C_{2,\mp}D_{-\frac{1}{2}+i\frac{\kappa_\mp^2}{2}}((-1+i)z_{\mp})\,,
\Eeq
where $C_{1,\mp}$ and $C_{2,\mp}$ are integration constants.

We determine the $C_{1,-}$ and $C_{2,-}$ constants by assuming that at early times, $-z_-\gg\max(1,|\kappa_-|)$, i.e., at $z_-\rightarrow-\infty$, the $\chi$ particles are in the adiabatic (i.e., WKB) vacuum,
\Beq
X_{\kappa_-}(z_-\rightarrow-\infty)=\frac{1}{\lambda^{1/4}\left|\dot{\bar{\phi}}_{\star,-}\right|^{1/2}}\frac{|z_-|^{-i\kappa_-^2/2}e^{-iz_-^2/2}}{\sqrt{2\sqrt{\kappa_-^2+z_-^2}}}\,.
\Eeq
By matching the asymptotic expansions of the parabolic cylinder functions at $z_-\rightarrow-\infty$ with this adiabatic vacuum solution we determine $C_{1,-}$ and $C_{2,-}$. 
Their explicit expressions as well as the asymptotic expansions of the parabolic cylinder functions and the WKB solution are given in the Supplemental Material.

\begin{figure}[t]
\includegraphics[width=3.3in]{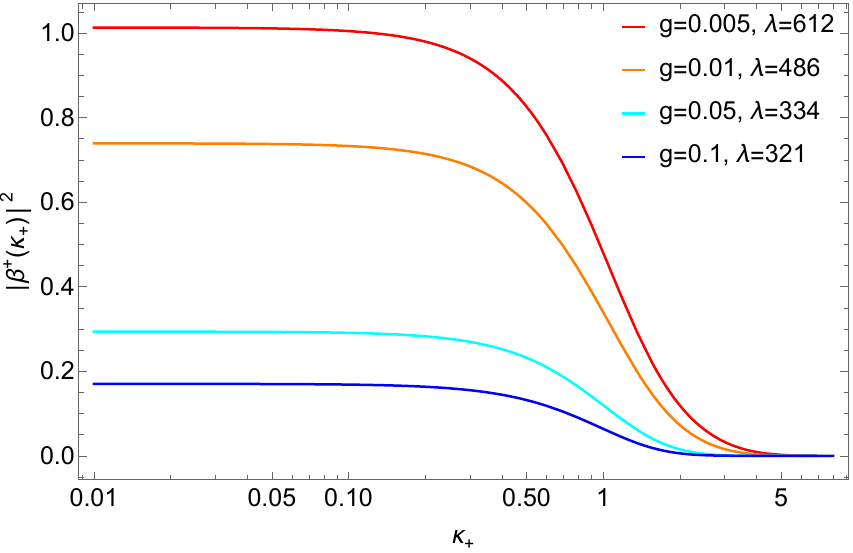}
\caption{The Bogoliubov coefficients for various values of the model parameter $g$. We observe sizable particle production for small wavenumbers, $\kappa_+\rightarrow0$, since only such modes can be non-adiabatically excited, see eq. \eqref{eq:nonad}; note that the values of $\beta$ are scale-independent since the adiabaticity is violated in a momentum-independent fashion. For large momenta $\beta$ decays exponentially fast, since on such scales adiabaticity is never violated and no particle production is expected. $\beta$ increases with decreasing $g$, since small $g$ correspond to efficient backreaction, implying more daughter particles.}
    \label{fig:beta}
\end{figure} 

To obtain $C_{1,+}$ and $C_{2,+}$, we require that the solution $X_k$
and its derivative $\dot X_k$ are continuous at $t=t_\star$, 
%This implies that at $z_{\mp}\rightarrow\mp0$ we have the two matching conditions
\Beq
&X_{\kappa_-(k)}(-0)=X_{\kappa_+(k)}(+0)\,,\\
&\left(\frac{d^2}{dz_-^2}X_{\kappa_-(k)}\right)(-0)=g\left(\frac{d^2}{dz_+^2}X_{\kappa_+(k)}\right)(+0)\,,
\Eeq
where the arguments $-0$ and $+0$ mean $z_{-}\to -0$ and $z_{+}\to+0$, respectively.
%We note that to do that we need the expressions for the parabolic cylinder functions and their derivatives at $z_{\mp}\rightarrow\mp0$. 
The explicit expressions for $C_{1,+}$ and $C_{2,+}$ are given in the Supplemental Material.

At late times, $z_+\rightarrow+\infty$, the solution to eq. \eqref{eq:Xpleom} relaxes into the WKB form,
\Beq
\label{eq:XplWKB}
X_{\kappa_+}&(z_+\rightarrow+\infty)=\frac{1}{\lambda^{1/4}\left|\dot{\bar{\phi}}_{\star,-}\right|^{1/2}}\times\\&\Bigg[\alpha^+_{\kappa_+}\frac{|z_+|^{-i\kappa_+^2/2}e^{-iz_+^2/2}}{\sqrt{2\sqrt{\kappa_+^2+z_+^2}}}
+\beta^+_{\kappa_+}\frac{|z_+|^{i\kappa_+^2/2}e^{iz_+^2/2}}{\sqrt{2\sqrt{\kappa_+^2+z_+^2}}}\Bigg]\,,
\Eeq
where $|\alpha^+|^2-|\beta^+|^2=1$.
By matching the asymptotic expansions of the parabolic cylinder functions at $z_+\rightarrow+\infty$ with this solution we determine the Bogoliubov coefficients, $\alpha^+$ and $\beta^+$. 
We note that $\beta^+\neq0$, implying that excitation of $\chi$-modes at $\phi=\phi_{\star}$ takes place, which is a manifestation of $\chi$-particle production.
The explicit expressions for $\beta^+$, which determines $\langle\chi^2\rangle(\phi)$, is given in the Supplemental Material. 

 In Fig. \ref{fig:beta}, we show $\beta^+(\kappa_+)$ for various model parameters. We note that $\beta^+(\kappa_+\rightarrow0)$ is $\kappa_+$-independent. On the other hand, $\beta^+(\kappa_+\rightarrow+\infty)$ decays exponentially fast,
since the violation of adiabaticity is exponentially suppressed at large $\kappa_+$.

{\it Effective inflaton potential}.--Given the amount of produced $\chi$-particles, see Fig. \ref{fig:beta}, we can calculate the effective inflaton potential.
In particular, we determine $\langle\hat{\chi}^2\rangle$ in eq.~\eqref{eq:Veff} using $\beta^+$,
%\Beq
%\label{eq:Vint2}
%V_{int}(\phi,\langle\chi^2\rangle)=\frac{\lambda}{2}(\phi-\phi_{\star})^2\langle\hat{\chi}^2\rangle\,,
%\Eeq
%After taking the positive frequency term from eq.~\eqref{eq:Xpleom}, we obtain
\Beq
\langle\hat{\chi}^2\rangle(t)=\int \frac{d^3k}{(2\pi)^3}\frac{|\beta^+_{\kappa_+(k)}|^2}{2a^3(t)\sqrt{\kappa_+^2(k)+z_+^2(t)}\lambda^{1/4}\left|\dot{\bar{\phi}}_{\star,-}\right|^{1/2}}\,.
\Eeq
Since $z_+\rightarrow+\infty$ and given that $\beta^+$ is sizable only for $\kappa_+<1$, we can ignore the $\kappa_+^2(k)$ term in the denominator. Hence, the final expression for $\langle\hat{\chi}^2\rangle$ is
%$\langle\hat{\chi}^2\rangle=0$ for $\phi>\phi_{\star}$ and
%\Beq
%\label{eq:chisq}
%&\rm{for}\quad \bar{\phi}<\phi_{\star}:\\
%\langle\hat{\chi}^2\rangle(t)
%=\begin{cases}
%0 &\mbox{for}~\phi>\phi_{\star}\,,
%\\
%\dfrac{\displaystyle\int \dfrac{d^3k}{(2\pi)^3}|\beta^+_{\kappa_+(k)}|^2}
%{2a^3(t)|z_+(t)|\lambda^{1/4}\left|\dot{\bar{\phi}}_{\star,-}\right|^{1/2}}~&\mbox{for}~\phi<\phi_{\star}\,.
%\end{cases}
%\Eeq
%and
%\Beq
%\label{eq:chi2}
%&\rm{for}\quad \bar{\phi}\geq\phi_{\star}:\quad\langle\hat{\chi}^2\rangle=0\,,
%\Eeq
%since no particle production takes place before $\bar{\phi}$ reaches $\phi_{\star}$. 

\begin{equation}\label{eq:chisq}
    \langle\hat{\chi}^2\rangle(t)=\dfrac{\displaystyle\int \dfrac{d^3k}{(2\pi)^3}|\beta^+_{\kappa_+(k)}|^2}
{2a^3(t)|z_+(t)|\lambda^{1/4}\left|\dot{\bar{\phi}}_{\star,-}\right|^{1/2}},
\end{equation}
for $\phi<\phi_{\star}$ and 0 elsewhere. We note that the integral in eq.~\eqref{eq:chisq} is time-independent.

In this effective scheme, the single inflaton field $\phi$ plays the role of a clock field. In particular,
\Beq
t-t_\star\approx-\frac{\bar{\phi}(t)-\phi_{\star}}{\left|\dot{\bar{\phi}}_{\star,+}\right|} \quad\rm{for}\quad \phi<\phi_{\star}\,, 
\Eeq
implying that $a$ and $z_+$ are functions of $\phi$. Specifically we have 
\Beq
&a(\phi)\approx a_\star e^{H_\star(t-t_\star)}\approx a_\star{\rm exp}\left(-H_\star\frac{(\phi-\phi_{\star})}{\left|\dot{\bar{\phi}}_{\star,+}\right|}\right)\,,\\
&z_+(\phi)\approx\lambda^{1/4}\frac{(\phi_{\star}-\phi)}{\left|\dot{\bar{\phi}}_{\star,+}\right|^{1/2}}\qquad\mbox{for}~\phi<\phi_{\star}\,,
\Eeq
where $H_\star=H(t_\star)$.

%We can do the following if we want a whole page figure:
\begin{figure*}[t]
    \includegraphics[width=0.8\textwidth]{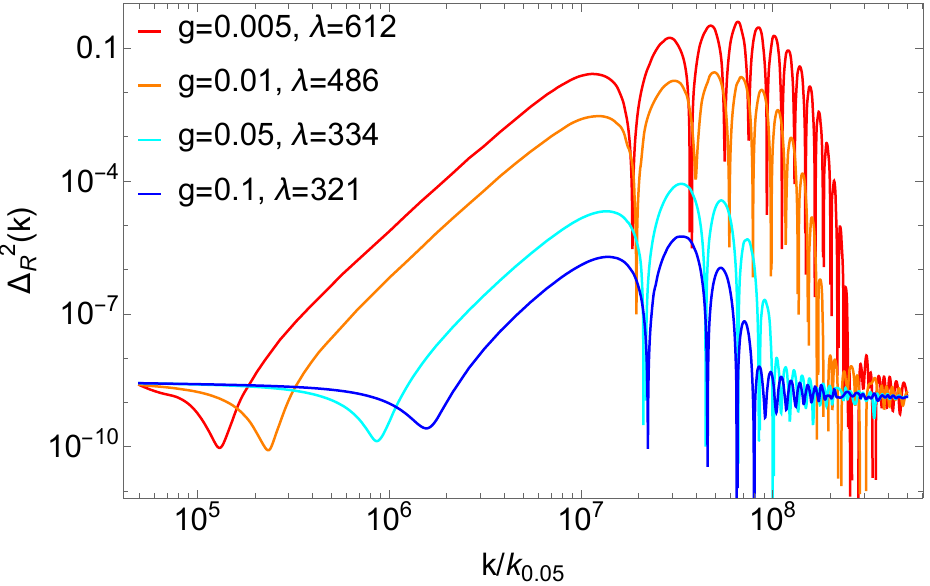}
    \caption{The curvature perturbation for $\phi_{\star}=3.38M$, $M=m_{pl}$, and various values of $\lambda$ and $g$, where $a(t_\star)H(t_\star)\approx5\times10^6\text{Mpc}^{-1}$.
    Modes which left the horizon around $\phi=\phi_{\star}$ have an enhanced spectrum which has a decaying oscillatory pattern. Modes which left the Hubble horizon when $\phi$ is far from $\phi_{\star}$ have the standard nearly-scale invariant power-spectrum. The pivot scale used for normalization in the horizontal axis is $k_{0.05}=0.05 {\rm Mpc}^{-1}$.
    Hence $a(t_\star)H(t_\star)/k_{0.05}\approx10^8$.
    }
    \label{fig:zetaps}
\end{figure*}

%\begin{figure}[t!]
%\includegraphics[width=3.3in]{FinalPzeta.pdf}
%\caption{The curvature perturbation for $\phi_{\star}=3.38M$, $M=m_{pl}$%, and $\lambda=0$, $1370$, $22000$, $97700$, $465000$ and $3893000$ corresponding to $g=1$, $0.5$, $0.1$, $0.05$, $0.025$ and $0.01$, respectively. Modes which left the horizon around $\phi=\phi_{\star}$ have an enhanced spectrum which has a decaying oscillatory pattern. Modes which left the Hubble horizon when $\phi\neq\phi_{\star}$ have the standard nearly-scale invariant power-spectrum. The pivot scale used for normalization in the horizontal axis is $k_{0.05}=0.05 {\rm Mpc}^{-1}$.
%}
%    \label{fig:zetaps}
%\end{figure} 

After substituting these expressions in eq.~\eqref{eq:chisq}, we finally obtain the effective interaction potential,
%\begin{align}\label{eq:Vintf}
%&{\rm for}\quad \phi<\phi_{\star}:\\
%&V_{int}(\phi,\langle\chi^2\rangle(\phi))={\rm exp}\left(\frac{3H_\star(\phi-\phi_{\star})}{g^{1/2}|\dot{\bar{\phi}}_{\star,-}|}\right)\\
%&\times\left[\frac{3H_\star|\phi_{\star}-\phi|}{g^{1/2}|\dot{\bar{\phi}}_{\star,-}|}-1\right]g^{3/4}\left|\dot{\bar{\phi}}_{\star,-}\right|^{3/2}\frac{\lambda^{5/4}}{8\pi^2}\int d\kappa_+\kappa_+^2|\beta^+|^2\,,
%&V_{int}(\phi,\langle\chi^2\rangle(\phi))\nonumber\\
%&~=\left\{
%\begin{array}{lll}
%\exp\left(\dfrac{3H_\star(\phi-\phi_{\star})}{g|\dot{\bar{\phi}}_{\star,-}|}\right)
%g^{3/2}\left|\dot{\bar{\phi}}_{\star,-}\right|^{3/2}&\\
%\qquad\times(\phi_{\star}-\phi)\dfrac{\lambda^{5/4}}{8\pi^2}\int d\kappa_+\kappa_+^2|\beta^+|^2
%&\mbox{for}~\phi<\phi_\star\,,\\
%~\\
%~0&\mbox{for}~\phi>\phi_\star\,.
%\end{array}
%\right.\nonumber\\
%\end{align}
\begin{equation}\label{eq:Vintf}
 \begin{split}
    V_{int}(\phi,\langle\chi^2\rangle(\phi))\nonumber=\exp\left(\dfrac{3H_\star(\phi-\phi_{\star})}{g|\dot{\bar{\phi}}_{\star,-}|}\right)
g^{3/2}\left|\dot{\bar{\phi}}_{\star,-}\right|^{3/2}&\\
\qquad\times(\phi_{\star}-\phi)\dfrac{\lambda^{5/4}}{8\pi^2}\int d\kappa_+\kappa_+^2|\beta^+|^2,
 \end{split}
\end{equation}
for $\phi<\phi_{\star}$ and 0 elsewhere.

Note that 
%he integral in eq.~\eqref{eq:Vintf} and $\dot{\bar{\phi}}_{\star,-}$ are constants, where the latter 
$\dot{\bar{\phi}}_{\star,-}$ is determined by the slow-roll equation.
The $|\beta^+|^2$-integral depends on the value of the parameter $g$, which should be self-consistently determined. Namely, the integral depends on the sharpness of the bump or $g$, which itself depends on the amount of particle production.
The factors involving $\dot{\bar{\phi}}_{\star,+}=g\dot{\bar{\phi}}_{\star,-}$ come from the time-dependence in the expression for $\langle\chi^2\rangle$ in eq.~\eqref{eq:chisq}, while those involving $\phi-\phi_\star$ 
%The $\dot{\bar{\phi}}_{\star,-}$ factors are everywhere multiplied by $g$, so the dependence is actually on $\dot{\bar{\phi}}_{\star,+}$.
%The interpretation is that produced particles and 
 come from the subsequent evolution.
 %which is determined by $\dot{\bar{\phi}}_{\star,+}$. 
%For $\dot{\bar{\phi}}_{\star,-}$ one can use a numerically obtained value or the approximate $|\dot{\bar{\phi}}_{\star,-}|=m_{pl}H_\star\sqrt{2\epsilon}\approx m_{pl}H_\star\sqrt{2\epsilon_V}$, where $\epsilon_V=(m_{pl}^2/2)(V'(\phi_{\star})/V(\phi_{\star}))^2$. 
For given values of
$\dot{\bar{\phi}}_{\star,-}$ and $\lambda$, the interaction potential gives rise to a bump in the inflaton potential shown in Fig.~\ref{fig:papV}, with $g$ determined self-consistently. 
However, for the sake of technical simplicity as well as for comparison with the previous literature on bumpy inflation, we first fix the value of $g$, then determine $\lambda$ from our numerical background evolution computations.
%%%%%%%%%%% MS: Aug 4
%We find that many of the terms appearing throughout the section follow scaling relations, simplifying our equations significantly. Namely, we find the following
%\xp{XP: I will double check these relations}
%\begin{equation}
%    \lambda\propto g^{-2.32},\hspace{5mm}\int \text{d}\kappa_+\kappa_+^2|\beta^+|^2\propto g^{-3/4}.
%\end{equation}
%Then, we may write 
%\begin{equation}
%    V_{int}(\phi,\langle\chi^2\rangle(\phi))\propto g^{-2.9}.
%\end{equation}

{\it Curvature enhancement}.--Knowing the effective inflaton potential, given by eq.~\eqref{eq:Veff} with eqs.~\eqref{eq:V} and the interaction potential, we can determine the evolution of the background by solving eqs.~\eqref{eq:KGeom} and \eqref{eq:Freom}.
We solve them by fixing $g$ and giving a test value for $\lambda$ first, and determine $\lambda$ iteratively until the computed value of $g$ agrees with the assumed value of $g$.
%by solving eqs.~\eqref{eq:KGeom} and \eqref{eq:Freom}. 
%From the solutions we compare the obtained value for $g$ with the chosen value of $g$. If the two do not agree, we change the value of $\lambda$ and solve eqs. \eqref{eq:KGeom} again for the same value of $g$. 
%We repeat this procedure until we find a value for $\lambda$ for which the assumed value of $g$ agrees with the one extracted from the background solution.
We can then use this background solution to determine the curvature perturbation.

To this end, we work in perturbed spatially-flat FLRW spacetime,
\begin{equation}
\begin{aligned}
    ds^2=N^2dt^2-\gamma_{ij}(dx^i+N^idt)(dx^j+N^jdt)\,.
\end{aligned}
\end{equation}
%with a lapse close to unity, $|N^2(t,x^k)-1|\ll1$, and a small shift, $|N^i(t,x^k)|\ll1$. 
After choosing co-moving slicing,
\begin{equation}
    \begin{aligned}
        \phi(t,x^i)=\bar{\phi}(t)\,,
    \end{aligned}
\end{equation}
and setting the anisotropic components of $\gamma_{ij}$ to zero, 
the gauge-invariant comoving curvature perturbation, $\mathcal{R}$, is given by the perturbation to the diagonal part of the induced metric,
\begin{equation}
    \gamma_{ij}(t,x^k)=a^2(t)e^{2\mathcal{R}(t,x^k)}\delta_{ij}\,.
\end{equation}
The second order action for $\mathcal{R}$ is \cite{Abolhasani:2019cqw}
\Beq
S_\mathcal{R}^{(2)}=m_{pl}^2\int dtd^3xa^3\epsilon\left[\dot{\mathcal{R}}^2-\frac{(\partial_i\mathcal{R})^2}{a^2}\right]\,.
\Eeq
The quantized Fourier modes of the curvature perturbation are
\begin{equation}
    \begin{aligned}
        &\hat{\mathcal{R}}(t,x^i)=\int \frac{d^3k}{(2\pi)^3}\left[e^{ik^jx^j}\mathcal{R}_k(t)\hat{c}_k+h.c.\right]\,,
    \end{aligned}
\end{equation}
where the time-independent $\hat{c}_k$ and $\hat{c}_k^\dagger$ operators play the role of annihilation and creation operators, respectively, and obey the commutation relation $[\hat{c}_k,\hat{c}_{k'}^\dagger]=\delta(k-k')$. The mode function obeys the classical equation of motion,
\Beq
\ddot{\mathcal{R}}_k+\left(3H+\frac{\dot{\epsilon}}{\epsilon}\right)\dot{\mathcal{R}}_k+\frac{k^2}{a^2}\mathcal{R}_k=0\,.
\Eeq
The background solutions determine its time-dependent coefficients, $H,\, \epsilon,\, \dot{\epsilon},\, a$. We solve the equation for each mode $k$ from the time when it is sub-horizon, $k\gg aH$ and is in its adiabatic vacuum,
\Beq
\mathcal{R}_{k\gg aH}=\frac{e^{-i\int (k/a)dt}}{\sqrt{4m_{pl}^2a^2\epsilon k}}\,.
\Eeq
The late time solutions, $k\ll aH$, i.e., $t\rightarrow\infty$ are used to determine the dimensionless curvature power spectrum,
\Beq
\Delta_\mathcal{R}^2=\frac{k^3}{2\pi^2}|\mathcal{R}_k(t\rightarrow\infty)|^2\,.
\Eeq

In Fig.~\ref{fig:zetaps}, we show the curvature power spectrum for different values of the parameter $g$ (corresponding to different values of $\lambda$). We see that co-moving modes which left the horizon when $\phi$ is far from $\phi_{\star}$ have a scale-invariant power-spectrum. On the other hand, modes which crossed out the Hubble horizon when $\phi\approx\phi_{\star}$ have an enhanced oscillatory power spectrum. For the values of $g$ that we considered, we find that the range of enhanced scales span roughly 2-4 orders of magnitude.
%Assuming our approximations are valid for $g\sim10^{-8}$ 
As shown in Fig.~\ref{fig:zetaps}, the curvature power spectrum enhancement depends sensitively on the parameter $g$ (although arguably less sensitively than to ad hoc inflaton potential parametrisations \cite{Cole:2023wyx} (see also \cite{Iovino:2025tcv,Profumo:2026qpn})), and hence the amount of $\chi$-particle backreaction. Lower values of $g$ correspond to sharper, larger drops in velocity around $t=t_\star$, leading to a larger enhancement of power. Specifically, we find the relation $\Delta_\mathcal{R}^2|_\text{peak}\approx A_sg^{-4}/8$. Of particular relevance, we find that for $g=5\times10^{-3}$ ($\lambda=612$) we may realise a peak amplitude of $\Delta_\mathcal{R}^2\sim\mathcal{O}(10^{-1})$, which can give rise to sizable GWs and PBH production. We note that this dependence on $g$ is far more sensitive than in the step-like inflaton potential studied in \cite{Cai:2021zsp} (in which $\Delta_\mathcal{R}^2\propto\epsilon^{-1}g^{-2}$).

%\kl{The amplitude can be increased further by lowering $g$, since $\mathcal{A}\propto g^{-4}$.}\xp{XP: I didn't quite find this relation, I have tried to incorporate the idea into the text...}

Furthermore, particle production during inflation generically gives rise to large non-Gaussianities \cite{Enqvist:2005qu,Barnaby:2010sq,Yu:2025bcj}, which can be probed through cosmological observations \cite{Chudaykin:2025vdh,Jung:2025nss}. Such non-Gaussian modifications to the curvature perturbation probability distribution function may have large impacts on the corresponding PBH mass function \cite{Byrnes:2012yx,Young:2013oia,Gow:2022jfb,Ferrante:2022mui,Pi:2024lsu}, as well as the SIGW spectrum \cite{Adshead:2021hnm,Zeng:2025cer,Li:2025met}. A quantitative calculation of the PBH abundance and scalar-induced gravitational-wave spectrum in our scenario therefore requires the non-Gaussian statistics of the curvature perturbation, together with the corresponding nonlinear evolution. We leave this analysis to future work.

{\it Conclusions}.--We derived for the first time a bumpy feature in the inflaton potential from first-principles calculations. 
To do so, we considered a spectator field coupled to the inflaton and studied the effects of particle production of the spectator field on the dynamics of the inflaton background.
%and the generation of curvature perturbation during inflation. 
We showed that for a certain form of the coupling, the backreaction can induce a bump-like feature in the inflaton potential. 
This feature can lead to a considerable enhancement of the curvature perturbation on scales much smaller than the CMB scale, and can consequently lead to the generation of PBHs and SIGWs, which is of interest to future GW observatories. 

Finally, let us mention a potential issue in this scenario. It is that it requires a large coupling constant, $\lambda\gg1$ in our setting. 
This raises a suspicion that higher order quantum effects may significantly affect the scenario. 
Whether a large coupling constant is an unavoidable feature of our scenario, and to what extent higher-order quantum corrections must be included, are deferred to future work.

\acknowledgments
This work is supported in part by JSPS KAKENHI No.~24K00624. XP is supported by an STFC studentship. XP would like to thank IPMU for their hospitality whilst undertaking the majority of this project. 

\bibliography{ref}

%%%%%%%%%%%%%%%%%%%%%%%%%%
% Supplemental Material %
%%%%%%%%%%%%%%%%%%%%%%%%%%
\clearpage
\onecolumngrid
\begin{center}
   \textbf{\large SUPPLEMENTAL MATERIAL \\[.1cm] ``Bumpy inflation from explosive particle production''}\\[.2cm]
  \vspace{0.05in}
  {Kaloian D. Lozanov, Xavier Pritchard, Misao Sasaki}
\end{center}

\twocolumngrid
\setcounter{equation}{0}
\setcounter{figure}{0}
\setcounter{table}{0}
\setcounter{section}{0}
\setcounter{page}{1}
\makeatletter
\renewcommand{\theequation}{S\arabic{equation}}
\renewcommand{\thefigure}{S\arabic{figure}}
\renewcommand{\thetable}{S\arabic{table}}

\onecolumngrid
%%%%%%%%%%%%%%%%%%%%%%%%%%

\section{WKB solutions}

The WKB solutions to eq. \eqref{eq:Xeom} are given by
\Beq
\label{eq:adiabplus}
    X_{\kappa_\mp}(z_\mp)=\frac{1}{\lambda^{1/4}\left|\dot{\bar{\phi}}_{\star,\mp}\right|^{1/4}}\frac{1}{\sqrt{2\omega_\mp}}\left[\alpha_{\kappa_\mp}^{\mp}e^{-i\int \omega_\mp dz_\mp}+\beta^{\mp}e^{i\int \omega_{\kappa_\mp} dz_\mp}\right]\,,
\Eeq
and hold in the adiabatic regime. The $\alpha$s and $\beta$s are the standard Bogoliubov coefficients.

To obtain the WKB solutions in the main text, note that $\omega_\mp=\int(\kappa_\mp^2+z_\mp^2)^{1/2}dz_\mp\approx z_\mp^2/2+\ln(z_\mp^2)^{\kappa_\mp^2/4}$.

\section{Asymptotics of parabolic cylinder functions}
The parabolic cylinder functions for large values of the variable are
\begin{equation}
\begin{aligned}
\label{eq:AsymptCylPar}
    \lim_{z_-\rightarrow-\infty}D_{-\frac{1}{2}-\frac{i\kappa_-^2}{2}}((1+i)z_-)=
    i&e^{-\pi\kappa_-^2/2}\Bigg[\frac{(-1)^{3/8}}{2^{(1+i\kappa_-^2)/4}}e^{-iz_-^2/2}e^{-3\pi\kappa_-^2/8}z_-^{-(1+i\kappa_-^2)/2}\\ &+\frac{(-1)^{7/8}\sqrt{2\pi}}{2^{(1-i\kappa_-^2)/4}\Gamma\left((1+i\kappa_-^2)/2\right)}\frac{e^{iz_-^2/2}e^{-\pi\kappa_-^2/8}}{z_-^{(1-i\kappa_-^2)/2}}\Bigg]\,,\\
    \lim_{z_-\rightarrow-\infty}D_{-\frac{1}{2}+\frac{i\kappa_-^2}{2}}((-1+i)z_-)=
    i&e^{-\pi\kappa_-^2/2}\Bigg[\frac{(-1)^{9/8}}{2^{(1-i\kappa_-^2)/4}}\frac{e^{iz_-^2/2}e^{\pi\kappa_-^2/8}}{z_-^{(1-i\kappa_-^2)/2}}\Bigg]\,,\\
    \lim_{z_+\rightarrow\infty}D_{-\frac{1}{2}-\frac{i\kappa_+^2}{2}}((1+i)z_+)=-&\frac{(-1)^{7/8}}{2^{(1+i\kappa_+^2)/4}}\frac{e^{-iz_+^2/2}e^{\pi\kappa_+^2/8}}{z_+^{(1+i\kappa_+^2)/2}}\,,\\
    \lim_{z_+\rightarrow\infty}D_{-\frac{1}{2}+\frac{i\kappa_+^2}{2}}((-1+i)z_+)=-&\frac{(-1)^{5/8}}{2^{(1-i\kappa_+^2)/4}}\frac{e^{iz_+^2/2}e^{-3\pi\kappa_+^2/8}}{z_+^{(1-i\kappa_+^2)/2}}+\frac{(-1)^{1/8}\sqrt{2\pi}}{2^{(1+i\kappa_+^2)/4}\Gamma\left((1-i\kappa_+^2)/2\right)}\frac{e^{-iz_+^2/2}e^{-\pi\kappa_+^2/8}}{z_+^{(1+i\kappa_+^2)/2}}\,.
\end{aligned}
\end{equation}
The parabolic cylinder functions take the following forms near the origin
\begin{equation}
\label{eq:NearOriginPar}
    \begin{aligned}
    \lim_{z_\mp\rightarrow0_\mp}D_{-\frac{1}{2}-\frac{i\kappa_\mp^2}{2}}&((1+i)z_\mp)=\frac{\sqrt{\pi}}{2^{(1+i\kappa_\mp^2)/4}\Gamma((1+(1+i\kappa_\mp^2)/2)/2)}\,,\\
    \lim_{z_\mp\rightarrow0_\mp}\partial_{z_\mp}D_{-\frac{1}{2}-\frac{i\kappa_\mp^2}{2}}&((1+i)z_\mp)=\pm\frac{-e^{i\pi/4}\sqrt{\pi}}{2^{(-3+i\kappa_\mp^2)/4}\Gamma((1+i\kappa_\mp^2)/4)}\,,\\
    \lim_{z_\mp\rightarrow0_\mp}D_{-\frac{1}{2}+\frac{i\kappa_\mp^2}{2}}&((-1+i)z_\mp)=\frac{\sqrt{\pi}}{2^{(1-i\kappa_\mp^2)/4}\Gamma((1+(1-i\kappa_\mp^2)/2)/2)}\,,\\
    \lim_{z_\mp\rightarrow0_\mp}\partial_{z_\mp}D_{-\frac{1}{2}+\frac{i\kappa_\mp^2}{2}}&((-1+i)z_\mp)=\pm\frac{e^{-i\pi/4}\sqrt{\pi}}{2^{-(3+i\kappa_\mp^2)/4}\Gamma((1-i\kappa_\mp^2)/4)}\,.
    \end{aligned}
\end{equation}

\section{Constants in solutions involving parabolic cylinder function}
The four constants in eq. \eqref{eq:Xparcylfncns} are
\Beq
C_{1,-}=-(-1)^{5/8} 2^{\frac{1}{4} i \left(\kappa _-^2+i\right)} e^{\frac{3 \pi  \kappa _-^2}{8}}\,,
\Eeq

\begin{equation}
\begin{aligned}
    C_{2,-}=\frac{(-1)^{3/8} \sqrt{\pi } 2^{\frac{1}{4}+\frac{i \kappa _-^2}{4}} e^{\frac{\pi  \kappa _-^2}{8}}}{\Gamma \left(\frac{i
   \kappa _-^2}{2}+\frac{1}{2}\right)}\,,
\end{aligned}
\end{equation}

\begin{equation}
\begin{aligned}
&C_{1,+}=\\
&-\frac{(-1)^{5/8} \sqrt[4]{g} 2^{\frac{1}{2} i g \kappa _+^2+\frac{i \kappa _+^2}{4}-\frac{3}{4}} e^{\frac{3}{8} \pi 
   g \kappa _+^2} \Gamma \left(\frac{3}{4}-\frac{i \kappa _+^2}{4}\right) \Gamma \left(\frac{i \kappa
   _+^2}{4}+\frac{3}{4}\right) \Gamma \left(\frac{1}{4} \left(i \kappa _+^2+1\right)\right) \Gamma \left(\frac{1}{4} i
   g \kappa _+^2+\frac{1}{4}\right)}{\pi ^{3/2} \Gamma \left(\frac{1}{2} i g \kappa _+^2+\frac{1}{2}\right)
   \left(\csc \left(\frac{1}{4} \pi  \left(1+i \kappa _+^2\right)\right)+i \sec \left(\frac{1}{4} \left(\pi +i \pi  \kappa
   _+^2\right)\right)\right)}\\
   &-\frac{\sqrt[8]{-1} 2^{\frac{1}{2} i g \kappa _+^2+\frac{i \kappa _+^2}{4}-\frac{3}{4}}
   e^{\frac{3}{8} \pi  g \kappa _+^2} \Gamma \left(\frac{1}{4}-\frac{i \kappa _+^2}{4}\right) \Gamma \left(\frac{i
   \kappa _+^2}{4}+\frac{3}{4}\right) \Gamma \left(\frac{1}{4} \left(i \kappa _+^2+1\right)\right) \Gamma \left(\frac{1}{4} i
   g \kappa _+^2+\frac{3}{4}\right)}{\pi ^{3/2} \sqrt[4]{g} \Gamma \left(\frac{1}{2} i g \kappa
   _+^2+\frac{1}{2}\right) \left(\csc \left(\frac{1}{4} \pi  \left(1+i \kappa _+^2\right)\right)+i \sec \left(\frac{1}{4}
   \left(\pi +i \pi  \kappa _+^2\right)\right)\right)}\\
   &+\frac{(1+i) \sqrt[8]{-1} \sqrt[4]{g} 2^{-\frac{3}{4}+\frac{i \kappa
   _+^2}{4}} e^{\frac{1}{8} \pi  g \kappa _+^2} \Gamma \left(\frac{3}{4}-\frac{i \kappa _+^2}{4}\right) \Gamma
   \left(\frac{i \kappa _+^2}{4}+\frac{3}{4}\right) \Gamma \left(\frac{1}{4} \left(i \kappa _+^2+1\right)\right) \cosh
   \left(\frac{1}{2} \pi  g \kappa _+^2\right) \Gamma \left(\frac{1}{4}-\frac{1}{4} i g \kappa _+^2\right)}{\pi
   ^2 \left(\csc \left(\frac{1}{4} \pi  \left(1+i \kappa _+^2\right)\right)+i \sec \left(\frac{1}{4} \left(\pi +i \pi  \kappa
   _+^2\right)\right)\right)}\\
   &+\frac{(1+i) \sqrt[8]{-1} 2^{-\frac{3}{4}+\frac{i \kappa _+^2}{4}} e^{\frac{1}{8} \pi  g
   \kappa _+^2} \Gamma \left(\frac{1}{4}-\frac{i \kappa _+^2}{4}\right) \Gamma \left(\frac{i \kappa
   _+^2}{4}+\frac{3}{4}\right) \Gamma \left(\frac{1}{4} \left(i \kappa _+^2+1\right)\right) \cosh \left(\frac{1}{2} \pi 
   g \kappa _+^2\right) \Gamma \left(\frac{3}{4}-\frac{1}{4} i g \kappa _+^2\right)}{\pi ^2 \sqrt[4]{g}
   \left(\csc \left(\frac{1}{4} \pi  \left(1+i \kappa _+^2\right)\right)+i \sec \left(\frac{1}{4} \left(\pi +i \pi  \kappa
   _+^2\right)\right)\right)}\,,
\end{aligned}
\end{equation}

\begin{equation}
\begin{aligned}
&C_{2,+}=\\
&\frac{\sqrt[8]{-1} \sqrt[4]{g} 2^{\frac{1}{2} i g \kappa _+^2-\frac{i \kappa _+^2}{4}-\frac{3}{4}} e^{\frac{3}{8} \pi 
   g \kappa _+^2} \Gamma \left(\frac{1}{4}-\frac{i \kappa _+^2}{4}\right) \Gamma \left(\frac{3}{4}-\frac{i \kappa
   _+^2}{4}\right) \Gamma \left(\frac{i \kappa _+^2}{4}+\frac{3}{4}\right) \Gamma \left(\frac{1}{4} i g \kappa
   _+^2+\frac{1}{4}\right)}{\pi ^{3/2} \Gamma \left(\frac{1}{2} i g \kappa _+^2+\frac{1}{2}\right) \left(\csc
   \left(\frac{1}{4} \pi  \left(1+i \kappa _+^2\right)\right)+i \sec \left(\frac{1}{4} \left(\pi +i \pi  \kappa
   _+^2\right)\right)\right)}\\
   &+\frac{\sqrt[8]{-1} 2^{\frac{1}{2} i g \kappa _+^2-\frac{i \kappa _+^2}{4}-\frac{3}{4}}
   e^{\frac{3}{8} \pi  g \kappa _+^2} \Gamma \left(\frac{1}{4}-\frac{i \kappa _+^2}{4}\right) \Gamma
   \left(\frac{3}{4}-\frac{i \kappa _+^2}{4}\right) \Gamma \left(\frac{i \kappa _+^2}{4}+\frac{1}{4}\right) \Gamma
   \left(\frac{1}{4} i g \kappa _+^2+\frac{3}{4}\right)}{\pi ^{3/2} \sqrt[4]{g} \Gamma \left(\frac{1}{2} i g
   \kappa _+^2+\frac{1}{2}\right) \left(\csc \left(\frac{1}{4} \pi  \left(1+i \kappa _+^2\right)\right)+i \sec
   \left(\frac{1}{4} \left(\pi +i \pi  \kappa _+^2\right)\right)\right)}\\
   &-\frac{(1-i) \sqrt[8]{-1} \sqrt[4]{g}
   2^{-\frac{3}{4}-\frac{1}{4} i \kappa _+^2} e^{\frac{1}{8} \pi  g \kappa _+^2} \Gamma \left(\frac{1}{4}-\frac{i
   \kappa _+^2}{4}\right) \Gamma \left(\frac{3}{4}-\frac{i \kappa _+^2}{4}\right) \Gamma \left(\frac{i \kappa
   _+^2}{4}+\frac{3}{4}\right) \cosh \left(\frac{1}{2} \pi  g \kappa _+^2\right) \Gamma \left(\frac{1}{4}-\frac{1}{4} i
   g \kappa _+^2\right)}{\pi ^2 \left(\csc \left(\frac{1}{4} \pi  \left(1+i \kappa _+^2\right)\right)+i \sec
   \left(\frac{1}{4} \left(\pi +i \pi  \kappa _+^2\right)\right)\right)}\\
   &-\frac{(1+i) \sqrt[8]{-1} 2^{-\frac{3}{4}-\frac{1}{4}
   i \kappa _+^2} e^{\frac{1}{8} \pi  g \kappa _+^2} \Gamma \left(\frac{1}{4}-\frac{i \kappa _+^2}{4}\right) \Gamma
   \left(\frac{3}{4}-\frac{i \kappa _+^2}{4}\right) \Gamma \left(\frac{i \kappa _+^2}{4}+\frac{1}{4}\right) \cosh
   \left(\frac{1}{2} \pi  g \kappa _+^2\right) \Gamma \left(\frac{3}{4}-\frac{1}{4} i g \kappa _+^2\right)}{\pi
   ^2 \sqrt[4]{g} \left(\csc \left(\frac{1}{4} \pi  \left(1+i \kappa _+^2\right)\right)+i \sec \left(\frac{1}{4} \left(\pi +i
   \pi  \kappa _+^2\right)\right)\right)}\,,
\end{aligned}
\end{equation}

The Bogoliubov coefficient in eq. \eqref{eq:XplWKB} is
\begin{equation}
\begin{aligned}
&\beta^+(\kappa_+)=\\
&\frac{(1-i) \sqrt[4]{g} 2^{-1+\frac{1}{2} i g \kappa _+^2} e^{\frac{1}{8} \pi  \left(g-3\right) \kappa
   _+^2+\frac{1}{4} \pi  g \kappa _+^2} \Gamma \left(\frac{1}{4}-\frac{i \kappa _+^2}{4}\right) \Gamma
   \left(\frac{3}{4}-\frac{i \kappa _+^2}{4}\right) \Gamma \left(\frac{i \kappa _+^2}{4}+\frac{3}{4}\right) \Gamma
   \left(\frac{1}{4} i g \kappa _+^2+\frac{1}{4}\right)}{\pi ^{3/2} \Gamma \left(\frac{1}{2} i g \kappa
   _+^2+\frac{1}{2}\right) \left(\csc \left(\frac{1}{4} \pi  \left(1+i \kappa _+^2\right)\right)+i \sec \left(\frac{1}{4}
   \left(\pi +i \pi  \kappa _+^2\right)\right)\right)}\\
   &+\frac{(1-i) 2^{-1+\frac{1}{2} i g \kappa _+^2} e^{\frac{1}{8}
   \pi  \left(g-3\right) \kappa _+^2+\frac{1}{4} \pi  g \kappa _+^2} \Gamma \left(\frac{1}{4}-\frac{i \kappa
   _+^2}{4}\right) \Gamma \left(\frac{3}{4}-\frac{i \kappa _+^2}{4}\right) \Gamma \left(\frac{i \kappa
   _+^2}{4}+\frac{1}{4}\right) \Gamma \left(\frac{1}{4} i g \kappa _+^2+\frac{3}{4}\right)}{\pi ^{3/2} \sqrt[4]{g}
   \Gamma \left(\frac{1}{2} i g \kappa _+^2+\frac{1}{2}\right) \left(\csc \left(\frac{1}{4} \pi  \left(1+i \kappa
   _+^2\right)\right)+i \sec \left(\frac{1}{4} \left(\pi +i \pi  \kappa _+^2\right)\right)\right)}\\
   &+\frac{i \sqrt[4]{g}
   e^{\frac{1}{8} \pi  \left(g-3\right) \kappa _+^2} \Gamma \left(\frac{1}{4}-\frac{i \kappa _+^2}{4}\right) \Gamma
   \left(\frac{3}{4}-\frac{i \kappa _+^2}{4}\right) \Gamma \left(\frac{i \kappa _+^2}{4}+\frac{3}{4}\right) \cosh
   \left(\frac{1}{2} \pi  g \kappa _+^2\right) \Gamma \left(\frac{1}{4}-\frac{1}{4} i g \kappa _+^2\right)}{\pi
   ^2 \left(\csc \left(\frac{1}{4} \pi  \left(1+i \kappa _+^2\right)\right)+i \sec \left(\frac{1}{4} \left(\pi +i \pi  \kappa
   _+^2\right)\right)\right)}\\
   &-\frac{e^{\frac{1}{8} \pi  \left(g-3\right) \kappa _+^2} \Gamma \left(\frac{1}{4}-\frac{i
   \kappa _+^2}{4}\right) \Gamma \left(\frac{3}{4}-\frac{i \kappa _+^2}{4}\right) \Gamma \left(\frac{i \kappa
   _+^2}{4}+\frac{1}{4}\right) \cosh \left(\frac{1}{2} \pi  g \kappa _+^2\right) \Gamma \left(\frac{3}{4}-\frac{1}{4} i
   g \kappa _+^2\right)}{\pi ^2 \sqrt[4]{g} \left(\csc \left(\frac{1}{4} \pi  \left(1+i \kappa _+^2\right)\right)+i
   \sec \left(\frac{1}{4} \left(\pi +i \pi  \kappa _+^2\right)\right)\right)}\,.
\end{aligned}
\end{equation}

Taking the modulus squared, this equation vastly simplifies upon several uses of the duplication formula, leaving

%\begin{equation}
%\begin{split}
%|\beta^+(\kappa_+)|^2=&\frac{e^{\frac{\pi(\sqrt{g}-1)k_+^2}{4}}\big|\Gamma\big(\frac{1-ik_+^2}{2}\big)\big|^2\cosh^2\big(\frac{\pi k_+^2}{2}\big)}{2\pi^3}\times\\
%&\Bigg|\frac{(1-i)\pi e^{\frac{\pi\sqrt{g}k_+^2}{4}}}{\sqrt{2}}\bigg(\frac{g^{1/8}\Gamma\big(\frac{3+ik_+^2}{4}\big)}{\Gamma\big(\frac{3+i\sqrt{g}k_+^2}{4}\big)}+\frac{\Gamma\big(\frac{1+ik_+^2}{4}\big)}{g^{1/8}\Gamma\big(\frac{1+i\sqrt{g}k_+^2}{4}\big)}\bigg)\\
%&+\cosh\bigg(\frac{\pi\sqrt{g}k_+^2}{2}\bigg)\bigg(ig^{1/8}\Gamma\bigg(\frac{3+ik_+^2}{4}\bigg)\Gamma\bigg(\frac{1-i\sqrt{g}k_+^2}{4}\bigg)-\frac{\Gamma\bigg(\frac{1+ik_+^2}{4}\bigg)\Gamma\bigg(\frac{3-i\sqrt{g}k_+^2}{4}\bigg)}{g^{1/8}}\bigg)\Bigg|^2.\\
%\end{split}
%\end{equation}
%This can be simplified slightly further, to arrive at

\begin{equation}
    |\beta^+(\kappa_+)|^2=\frac{1}{4\pi}e^{-\frac{\pi k_+^2(1+g)}{4}}(1+\cosh(\pi k_+^2))\bigg| \Gamma\bigg(\frac{1-ik_+^2}{2}\bigg)\bigg(\frac{\Gamma\big(\frac{1+ik_+^2}{4}\big)}{g^{1/4}\Gamma\big(\frac{1+igk_+^2}{4}\big)}-\frac{g^{1/4}\Gamma\big(\frac{3+ik_+^2}{4}\big)}{\Gamma\big(\frac{3+igk_+^2}{4}\big)}\bigg) \bigg|^2\,.
\end{equation}

\end{document}